\documentstyle[sprocl]{article}

\input{psfig}

\def\be{\begin{equation}}
\def\ee{\end{equation}}
\def\bea{\begin{eqnarray}}
\def\eea{\end{eqnarray}}

\begin{document}

\title{IS OUR VACUUM STABLE ? \footnote{Based on a lecture given
by E. I. Guendelman in the symposium "The Future of the Universe
and the Future of our Civilization" July 2-6, 1999} }

\author{E. I. GUENDELMAN and J. PORTNOY}

\address{Physics Department, Ben Gurion University of the Negev,
Beer Sheva, 84105 Israel\\E-mail: guendel@bgumail.bgu.ac.il, jportnoy@
bgumail.bgu.ac.il}


\maketitle\abstracts{
The stability of our vacuum is analyzed and several aspects concerning
this question are reviewed. 1) In the standard Glashow-Weinberg-Salam
(GWS) model we review the instability towards the formation of a
bubble of lower energy density and how the rate of such bubble formation
process compares with the age of the Universe for the known values
of the GWS model. 2) We also review the recent work by one of us (E.I.G)
concerning the vacuum instability question in the context of a model
that solves the cosmological constant problem. It turns out that in such
model the same physics that solves the cosmological constant problem
makes the vacuum stable. 3) We review our recent work concerning the
instability of elementary particle embedded in our vacuum, towards the
formation of an infinite Universe. Such process is not catastrophic. 
It leads to a "bifurcation type" instability in which 
our Universe  is not eaten by a bubble (instead a baby universe
is born). This universe does not replace our Universe rather 
it disconnects from it (via a wormhole) after formation.}

\section{On Vacuum Stability in the Weinberg-Salam model.}
In modern theories of elementary particle interactions, the vacuum
is defined by the expectation value of a scalar field. In the
Weinberg-Salam\cite{P1} model such scalar field (the Higgs field) is
responsible for the gauge symmetry breaking and fermion masses production.
Let us begin our studies of vacuum instability by considering a model
containing just a scalar field with a potential $V(\phi)$ as depicted in
fig 1.\\

\begin{figure}
\begin{center}
\psfig{figure=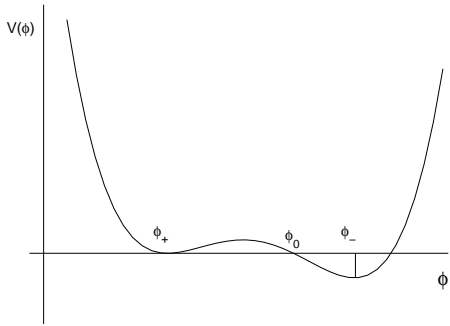,width=6.cm}
\caption{Scalar Field Potential with an Unstable Vacuum }
\end{center}
\end{figure}

The theory is governed by an action in the form
$S=\int d^{4}x (\frac{1}{2} \partial_{\mu} \phi \partial^{\mu} \phi - V(\phi))$,
where $\partial_{\mu} \phi \partial^{\mu} \phi = (\partial_{0})^{2} - (\nabla  \phi)^{2}$

As it is obvious, there is no classical (in Minkowski space) solution
connecting $\phi_{+}$ and $\phi_{-}$. There is however a Euclidean
solution connecting a point close to $\phi_{-}$ to $\phi_{+}$, the
tunneling solution. This is a solution in imaginary time. Defining $\tau = it$,
$\rho = \sqrt{\vec{x}^{2} + \rho^{2}}$ and considering solutions of the form
$\phi = \phi(\rho)$, we obtain the following equation of motion
\begin{equation}
\frac{d^{2} \phi}{d \rho^{2}} + \frac{3}{\rho} \frac{d \phi}{d \rho} + V^{'}(\phi) =0
\end{equation}
Equation (1) is like that of a particle, $\phi$ playing the role of
"position", $\rho$ "time", $\frac{3}{\rho} \frac{d \phi}{d \rho}$ friction,
$-V$, the mechanical potential (see$ -V$ in Fig.2).

\begin{center}
\begin{figure}

\psfig{figure=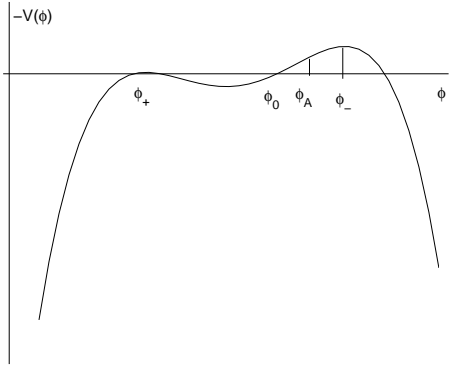,width=6.cm }
\caption{The Effective Mechanical Potential }
\end{figure}
\end{center}

To avoid a singularity at $\rho = 0$, $\frac{d \phi}{d \rho}$ must
vanish for $\rho = 0$. If we set as initial condition at $\rho = 0$,
$\phi$ close to $\phi = \phi_{-}$ the field stays close to $\phi = \phi_{-}$
for a long $\rho$ interval, the friction term becomes negligible
and $\phi$ arrives at $\phi = \phi_{+}$ with $\frac{d \phi}{d \rho} \neq 0$,
i.e. "overshoots". If $\phi = \phi_{0}$, because of the friction the
field does not make it to $\phi = \phi_{+}$. By continuity of a value
$\phi = \phi_{A}$ exists such that $\phi \rightarrow \phi_{+}$ as
$\rho \rightarrow \infty$. This is the tunneling solution.
This solution has a classical Euclidean action $S_{E}$ and the
tunneling probability per unit time per unit volume is proportional to
$e^{-S_{E}}$. $\phi_{A} \rightarrow \phi_{-}$ takes place classically.\\
In the quantum theory one finds zero-point fluctuations that give
rise to vacuum energy $\frac{1}{2} \hbar \omega$ for each boson mode
$(\omega = \sqrt{\vec{p}^{2} + m^{2}})$, since masses depend on the Higgs
field ($m^{2} = V^{"}(\phi)$ for the Higgs field itself, $m^{2} \sim e^{2} \phi^{2}$
for gauge vector fields, etc.) and $\frac{1}{2} \Sigma \hbar \omega
\rightarrow \frac{1}{2} V \int \frac{d^{3}k}{(2 \pi)^{3}} \hbar \sqrt{\vec{k}^{2} + m^{2}
(\phi)}$, we see that an infinite $\phi$ dependent vacuum energy appears.
The Dirac fermions also contribute $ - \hbar \omega $ for each occupied
negative energy state.
The resulting energy density $V_{eff}(\phi)$ makes sense
 after a renormalization process which get rid of the infinities.
The process of renormalization introduces an arbitrary scale into the
problem. If we demand that such scale will not affect physical quantities
like $V_{eff}(\phi)$ itself, we arrive at what is called renormalization
group equation for $V_{eff}(\phi)$.
Such  $V_{eff}(\phi)$ holds for small coupling although $\phi$ itself
can be large. \\       
One can then ask: for which values $V_{eff}(\phi)$ our
vacuum is unstable?               
One can check that for the known values of $m_{Higgs}>90Gev$,$V_{eff}(\phi=0)
> V_{eff}(\phi_{+})$ ($\phi_{+}$ is the value of the Higgs for our vacuum),
but for large values of $\phi$, $V_{eff}(\phi)$ can be negative
(conventionally we set $V_{eff}(\phi_{+}) \equiv 0$). The condition
that does not happen gives\cite{P2} $m_{top} \leq 95 Gev + 0.60 m_{Higgs}$

Notice that for $m_{top} \approx 175 Gev$, this implies that
 $m_{Higgs} \geq 135Gev$,
 while the experimental data only tell us that
$95Gev \leq m_{Higgs} \leq 190Gev$. So that we cannot tell for sure
whether the standard model predicts a stable or unstable vacuum.\\
Even assuming the vacuum is unstable, which is possible according
to what we have seen before, can ask: was the age of the universe long
enough so that the probability of some nucleation took
place in our past light cone?
The space-time volume of our past light cone $\sim t^{4}$, $t$ is
the age of the universe and the probability of nucleating per
unit time per unit volume $\sim e^{-S_{E}}$. The probability of
nucleation is  $\sim t^{4} e^{-S_{E}}$ and the age of the universe
in electroweak units is $t \sim e^{101}$ so that $t^{4} e^{-S_{E}}
\sim e^{-S_{E}+404}$. So that $S_{E} \geq 404$ gives the middle region  Fig. 3
where the rate is not high enough so it is unlikely that we noticed already
any phase transition.\\
The value $S_{E} \geq 404$ is consistent with the known bounds on
$m_{H}$ and the known  value of $m_{top}$ (See Fig. 3 which contains
the inserted horizontal line $m_{top} =175 Gev, 90 Gev < m_{H} < 190 Gev $) .

\begin{center}
\begin{figure}

\psfig{figure=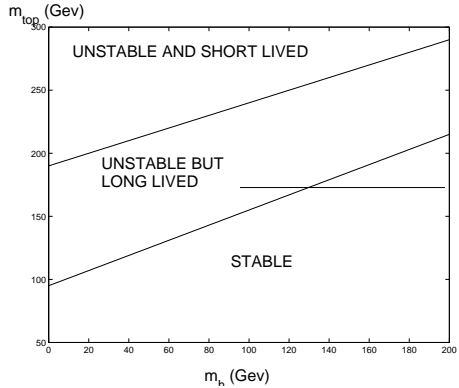,width=6.cm}
\caption{Regions of Stability and Unstability }
\end{figure}
\end{center}

\section{The Cosmological Constant Problem}

Calculations of $V_{eff}(\phi)$ are generally  a calculation
of the vacuum energy of the universe and how it depends on $\phi$.
In flat space there is no meaning to the constant part of $V_{eff}(\phi)$
if gravity is ignored. Once gravity taken into account is very
important. The naive prediction for $V_{eff}(\phi)|_{\phi=\phi_{min}}=\infty$
but we really need $V_{eff}(\phi)|_{\phi=\phi_{min}}=0$.

One has the  suspicion that the calculations of $V_{eff}(\phi)$
mentioned above may miss some physics. Fine tuning $V_{eff}(\phi)$,
by adding a constant so that $V_{eff}(\phi)=0$ may not be enough
as we will argue, since the physics that sets $V_{eff}(\phi_{+})\approx 0$
can act again when $V_{eff}$ "tries" to cross once again the zero value
and this could affect seriously our notions concerning stability of
the vacuum.

\section{A Model that sets $V_{eff}(\phi_{+})=0$ naturally}

Observation: Usually Generally Covariant theory is build from
the action $S=\int L \sqrt{-g} d^{4}x$, $g=det(g_{\mu \nu})$, $L=scalar$,
($L'=L $ under general coordinate transformations). Notice that
$\sqrt{-g} d^{4}x$ is an invariant volume element.
It is possible to build another volume element, independent of
$g_{\mu \nu}$. Take  for example, given 4-scalars $\varphi_{a}$ (a =
1,2,3,4), the density

\begin{equation}
\Phi =  \varepsilon^{\mu\nu\alpha\beta}  \varepsilon_{abcd}
\partial_{\mu} \varphi_{a} \partial_{\nu} \varphi_{b} \partial_{\alpha}
\varphi_{c} \partial_{\beta} \varphi_{d}
\end{equation}

$\Phi$ transform like $\sqrt{-g}$, so  $\Phi d^{4}x$ is also an invariant.

One can allow both geometrical
objects to enter the theory and consider\cite{P3}

\begin{equation}
S = \int L_{1} \Phi  d^{4} x  +  \int L_{2} \sqrt{-g}d^{4}x
\end{equation}

Here $L_{1}$ and $L_{2}$ are
$\varphi_{a}$  independent. There is a good reason not to consider
mixing of  $\Phi$ and
$\sqrt{-g}$ , like
for example using
$\frac{\Phi^{2}}{\sqrt{-g}}$. This is because (3) is invariant (up to the integral of a total
divergence) under the infinite dimensional symmetry
$\varphi_{a} \rightarrow \varphi_{a}  +  f_{a} (L_{1})$
where $f_{a} (L_{1})$ is an arbitrary function of $L_{1}$ if $L_{1}$ and
$L_{2}$ are $\varphi_{a}$
independent. Such symmetry (up to the integral of a total divergence) is
absent if mixed terms are present.

We will study now the dynamics of a scalar field $\phi$ interacting
with gravity as given by the action (3) with\cite{P4}

\begin{equation}
L_{1} = \frac{-1}{\kappa} R(\Gamma, g) + \frac{1}{2} g^{\mu\nu}
\partial_{\mu} \phi \partial_{\nu} \phi - V(\phi),  L_{2} = U(\phi)
\end{equation}

\begin{equation}
R(\Gamma,g) =  g^{\mu\nu}  R_{\mu\nu} (\Gamma) , R_{\mu\nu}
(\Gamma) = R^{\lambda}_{\mu\nu\lambda}, R^{\lambda}_{\mu\nu\sigma} (\Gamma) = \Gamma^{\lambda}_
{\mu\nu,\sigma} - \Gamma^{\lambda}_{\mu\sigma,\nu} +
\Gamma^{\lambda}_{\alpha\sigma}  \Gamma^{\alpha}_{\mu\nu} -
\Gamma^{\lambda}_{\alpha\nu} \Gamma^{\alpha}_{\mu\sigma}.
\end{equation}

In the variational principle $\Gamma^{\lambda}_{\mu\nu},
g_{\mu\nu}$, the scalar fields 
$\varphi_{a}$ and the  scalar field $\phi$ are to be treated
as independent variables.

We can require the scale invariance of the theory.
If we perform the global scale transformation ($\theta$ =
constant) $g_{\mu\nu}  \rightarrow   e^{\theta}  g_{\mu\nu}$
then (2), with the definitions (3), (4), is invariant.  $V(\phi)$
and $U(\phi)$ are in the form
$V(\phi) = f_{1}  e^{\alpha\phi},  U(\phi) =  f_{2}
 e^{2\alpha\phi}$ and $\varphi_{a}$ is transformed according to
$\varphi_{a}   \rightarrow   \lambda_{a} \varphi_{a}$
(no sum on a) which means
$\Phi \rightarrow \biggl(\prod_{a} {\lambda}_{a}\biggr) \Phi \equiv \lambda
\Phi $ such that $\lambda = e^{\theta}$ and
$\phi \rightarrow \phi - \frac{\theta}{\alpha}$. In this case we call the
scalar field $\phi$ needed to implement the scale invariance as "dilaton".

Now,in the general case, let us consider the equations which are obtained
from the variation of the $\varphi_{a}$ fields. We obtain then
$A^{\mu}_{a} \partial_{\mu} L_{1} = 0$
where  $A^{\mu}_{a} = \varepsilon^{\mu\nu\alpha\beta}
\varepsilon_{abcd} \partial_{\nu} \varphi_{b} \partial_{\alpha}
\varphi_{c} \partial_{\beta} \varphi_{d}$. Since
det $(A^{\mu}_{a}) =\frac{4^{-4}}{4!} \Phi^{3} \neq 0$ if $\Phi\neq 0$.
Therefore if $\Phi\neq 0$ we obtain that $\partial_{\mu} L_{1} = 0$, or that
$L_{1}  = M$, where M is constant. This constant M appears in a
self-consistency condition of the equations of motion
that allows us to solve for $ \chi \equiv \frac{\Phi}{\sqrt{-g}}$

\begin{equation}
\chi = \frac{2U(\phi)}{M+V(\phi)}.
\end{equation}

        To get the physical content of the theory, it is convenient to go
to the Einstein conformal frame where

\begin{equation}
\overline{g}_{\mu\nu} = \chi g_{\mu\nu}
\end{equation}

and $\chi$  given by (6). In terms of $\overline{g}_{\mu\nu}$   the non
Riemannian contribution (defined   as
$\Sigma^{\lambda}_{\mu\nu} =
\Gamma^{\lambda}_{\mu\nu} -\{^{\lambda}_{\mu\nu}\}$
where $\{^{\lambda}_{\mu\nu}\}$   is the Christoffel symbol),
disappears from the equations, which can be written then in the Einstein
form ($R_{\mu\nu} (\overline{g}_{\alpha\beta})$ =  usual Ricci tensor)

\begin{equation}
R_{\mu\nu} (\overline{g}_{\alpha\beta}) - \frac{1}{2}
\overline{g}_{\mu\nu}
R(\overline{g}_{\alpha\beta}) = \frac{\kappa}{2} T^{eff}_{\mu\nu}
(\phi)
\end{equation}

where

\begin{equation}
T^{eff}_{\mu\nu} (\phi) = \phi_{,\mu} \phi_{,\nu} - \frac{1}{2} \overline
{g}_{\mu\nu} \phi_{,\alpha} \phi_{,\beta} \overline{g}^{\alpha\beta}
+ \overline{g}_{\mu\nu} V_{eff} (\phi),
V_{eff} (\phi) = \frac{1}{4U(\phi)}  (V+M)^{2}.
\end{equation}

If $V(\phi) = f_{1} e^{\alpha\phi}$  and  $U(\phi) = f_{2}
e^{2\alpha\phi}$ as
required by scale invariance, we obtain from (10)
$V_{eff}  = \frac{1}{4f_{2}}  (f_{1}  +  M e^{-\alpha\phi})^{2}$

Since we can always perform the transformation $\phi \rightarrow
- \phi$ we can choose by convention $\alpha > O$.
We then see that as $\phi \rightarrow
\infty, V_{eff} \rightarrow \frac{f_{1}^{2}}{4f_{2}} =$ const.
providing an infinite flat region. Also a minimum is achieved at zero
cosmological constant for the case $\frac{f_{1}}{M} < O$ at the point
$\phi_{min}  =  \frac{-1}{\alpha} ln \mid\frac{f_1}{M}\mid $. Finally,
the second derivative of the potential  $V_{eff}$  at the minimum is
$V^{\prime\prime}_{eff} = \frac{\alpha^2}{2f_2} \mid{f_1}\mid^{2} > O$
if $f_{2} > O$,

There are many interesting issues that one can raise here. The first one
is of course the fact that a realistic scalar field potential, with
massive excitations when considering the true vacuum state, is achieved in
a way which is consistent with the idea of the scale invariance. The second point 
to be raised is that since there is an infinite
region of flat potential for $\phi \rightarrow \infty$, we expect a slow
rolling new
inflationary scenario to be
viable, provided the universe is started at a sufficiently large value of
the scalar field $\phi$.
Furthermore, one can consider this model as suitable for the
present day universe rather than for the early universe, after we suitably
reinterpret the meaning of the scalar field  $\phi$. This can provide a long
lived almost constant vacuum energy for a
long period of time, which can be small if $f_{1}^{2}/4f_{2}$ is
small.

Such small energy
density will eventually disappear when the universe achieves its true
vacuum state. \\
Notice that for generic functions $V(\phi)$, $U(\phi)$ the minimum of
$V_{eff}(\phi)$, as given from (9), is at zero if $V+M=0$ at some point
and if $V'$ is finite there (also $U > 0$ there).
So $V'_{eff} = 0$ and $V_{eff}=0$ is achieved generally
without fine tuning!
If in the neighborhood of $\phi_{+}$ $V + M = 0$, $U(\phi)>0$ and
$V + M $ as a function of $\phi$ that goes through zero.
Then $V_{eff} \sim \frac{(V+M)^{2}}{4 U(\phi_{+})}$ has a local
minimum at zero. That is $V(\phi_{+}) =0$ and $V(\phi_{+})=0$ automatically without
fine tuning. Therefore zero vacuum energy state is obtained naturally!\\
Going back to the general $V(\phi)$, $U(\phi)$ case we can ask the
question: given the classically stable state $\phi_{+}$, where
$\phi=\phi_{+}$, $U(\phi_{+})>0$, $V(\phi_{+})+M=0$ can we
make this into an unstable state?. Remember that
$V_{eff}=\frac{(V+M)^{2}}{4 U(\phi_{+})}$ and that we obtained
$V+M=0$ as a stable (classically) state under the conditions
that $U$ is $> 0$ at this point and it is a regular function there.
We take to be true that $U(\phi)$ is a nice function everywhere.
The only way $V_{eff}=\frac{(V+M)^{2}}{4 U}$ can change sign is for
$U(\phi)$ to change sign. For $U(\phi)$ being a nice function, this
can only happen if $U(\phi)$ goes to zero. If no fine tuning is
invoked at that point, $V+M \neq  0$ (if it is true, for other $M$,
i.e. for another initial condition of the universe it will not be true).
Then $V_{eff}$ looks like in Fig.4

\begin{center}
\begin{figure}

\psfig{figure=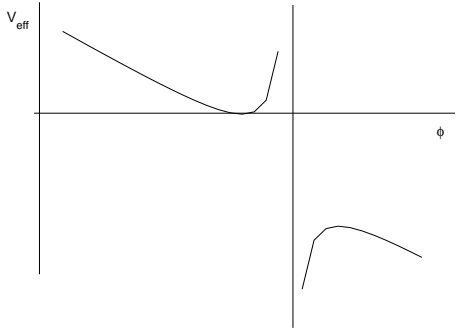,width=6.cm}
\caption{The Effective Potential for the Case $U(\phi)$ contains zero }
\end{figure}
\end{center}
Remember that in the euclidean solution relevant for nucleation
-$V_{eff}$ is the relevant potential.

\begin{center}
\begin{figure}

\psfig{figure=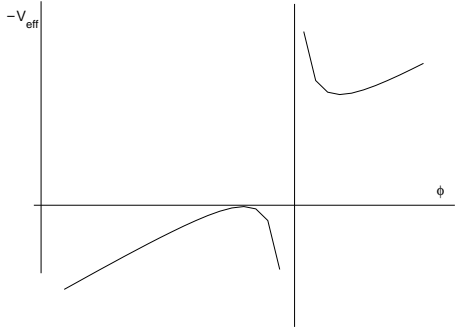,width=6.cm}
\caption{Effective Mechanical Potential when $U(\phi)$ contains  zero }
\end{figure}
\end{center}
It is clear from Fig.5 that no tunneling is possible now!\\
From eq.(6) we have $V+M \rightarrow 0 \Rightarrow |\Phi|>>\sqrt{-g}$ and
$U \rightarrow 0 \Rightarrow |\Phi|<<\sqrt{-g}$.\\
It is interesting to see what the volume element $\sqrt{-\bar{g}} d^{4}x$
where $\bar{g}_{\mu \nu} = \chi g_{\mu \nu}=$ Einstein metric compares with
the volume element $\Phi d^{4}x$. Indeed, $\sqrt{-\bar{g}}= \chi^{2} \sqrt{-g}
= \chi \Phi$, so that $\sqrt{-\bar{g}}>>|\Phi|$ for $V+M \rightarrow 0$,
i.e. Einstein energy density gets diluted $\Rightarrow V_{eff} \rightarrow 0$.
On other limit, $U \rightarrow 0$, $\sqrt{-\bar{g}}<<|\Phi|$, implies
that Einstein energy density gets concentrated.This is
the physical reason that leads to the energy barrier against
crossing from $V_{eff}>0$ to $V_{eff}<0$.
\section{The Decay of an Elementary Particle into a Universe}
Finally we consider a different type of instability of our vacuum.
In this case a region which contains a false vacuum inside, like
an elementary particle, can decay into a universe!\\
We considered a model of an elementary particle \cite{YG}
as a $ 2 + 1 $ dimensional brane evolving in a $ 3 + 1 $ dimensional
space. The introduction of a gauge field which takes place in the brane as well
as a  normal surface tension, following the standard approach
to the theory of extended objects \cite{N},
 can lead to a stable "elementary particle"
configuration. The simplest form of the action that permits  a stable
configuration is:
\begin{equation}
S=\sigma _{0} \int \sqrt{-h} d^{3}y + \lambda \int \sqrt{-h}
F_{\alpha \beta} F^{\alpha \beta} d^{3}y
\end{equation}
Where $h_{\alpha \beta}, \ \alpha \ \beta = 0, 1, 2 $
is the induced metric on the surface of the  membrane,
 $h=det(h_{\alpha \ \beta} ) $,  and
 $F_{\alpha \beta} F^{\alpha \beta}$ the Lagrangian
of the gauge field. If we assume a spherically symmetric
 vector potential in the brane (up to a gauge transformation)
of the simplest form (a monopole potential), we receive
the general form of the surface tension as;
$\sigma = \sigma _{0} + \frac {\sigma_{1}}{r^{4}}$
($\sigma_{1}$  being  $2 \lambda f^{2}$,  f being the strength
of the monopole configuration defined by the vector
potential in the brane).
The energy of the static wall, $4 \pi r^{2}\sigma$, has
a non trivial minimum for any $ \lambda > 0 $ that permits a stable
configuration.  This is the simplest possible model, below we shall
consider the effect of gravity  and of an internal vacuum energy.
A  membrane as discussed above, defines boundaries between different
phases with different values for their energy densities.

We took the metric for the inside of the membrane phase, a false vacuum
one, as a de Sitter metric, i.e., $ds^{2} = -(1-\chi ^{2} r^{2})dt^{2} +
\frac {dr^{2}} {(1-\chi ^{2} r^{2})} +
r^{2}(d\theta ^{2} + sin^{2} \theta d\phi ^{2})$
where $\chi ^{2} \ =\ \frac {8\pi \rho G}{3}$, $\rho$ is the energy density.
Outside, in the empty space, we can have only a Schwarzschild space
time according to Birkhoff's theorem, i.e.,
$ds^{2}= - (1 - \frac {2GM}{r})dt^{2} + \frac {dr^{2}}{(1 - \frac {2GM}{r})} +
r^{2}(d\theta ^{2} + sin^{2} \theta d\phi ^{2})$
and in the membrane, we have a singular energy momentum tensor.\\
Demanding that Einstein's equations to be satisfied not only inside
and outside but also in the membrane we get \cite {YG}
$\sqrt{1-\chi ^{2} r^{2}+\dot {r}^{2}}-\sqrt{1-\frac {2GM}{r}+\dot {r}^{2}}
= 4 \pi G \sigma r$
where $\dot {r} \equiv \frac {dr}{d \tau}$ and $\sigma$ is the one discussed
before.
 Following (7), we take as the Hamiltonian the mass of
the system, which gives us:
\begin{equation}
H=\frac{\chi ^{2} r^{3}}{2G} - \frac {(4\pi G)^{2} \sigma ^{2} r^{3}}{2G}
+4 \pi \sigma r^{2} (1-\chi ^{2} r^{2} + \dot{r} ^{2})^{1/2}
\end{equation}
Having obtained the Hamiltonian, all the others classical dynamical variables
can be obtained as was done in \cite {B}. The conjugate momentum p will
be equal to

$p=\frac{\partial L}{\partial \dot{r}}$,
the Lagrangian will be equal to
$L=\dot {r} \int H \frac{d\dot{r}}{\dot{r}^{2}}$
This give for the conjugate momentum
$p=\int \frac{\partial H}{\partial \dot{r}} \frac{d\dot{r}}{\dot{r}}$
Using H as before we arrive at the value of p, which is equal to
$p=4 \pi \sigma r^{2} arcsinh(\frac{\dot{r}}{\sqrt{1-\chi^{2} r^{2}}})$
 inside horizon and
$p=4 \pi \sigma r^{2} arccosh(\frac{\dot{r}}{\sqrt{\chi^{2} r^{2}-1}})$
outside.
An arbitrary function of r can be added in the definition of p. Classically
it corresponds to an additional total derivative of a function of r in
the Lagrangian, while in Quantum Mechanics it corresponds to a redefinition
of the wavefunction $\Psi '=e^{if(r)} \Psi$
This means that the Hamiltonian can be taken as
\begin{equation}
H=\frac{\chi ^{2} r^{3}}{2G} - \frac {(4\pi G)^{2} \sigma ^{2} r^{3}}{2G}
+4 \pi \sigma r^{2} \sqrt{1-\chi ^{2} r^{2}}K(\frac{p}{4 \pi \sigma r^{2}})
\end{equation}
where $K=cosh$ inside the horizon and $K=sinh$ outside it.
In order to achieve a quantum mechanical approach we shall assume that
$p=-i\frac{\partial}{\partial r}$
and from this
$e^{-ia\frac{\partial}{\partial r}}\Psi (r)=\Psi(r-ia)$.
 The Schroedinger equation is
$H\Psi=m\Psi$
in which m is the mass parameter of the external Schwarzschild  solution.
Defining the dimensionless variable (in units where $\hbar$ = c = 1)
$x=\frac{4\pi r^{3}\sigma _{0}}{3}-4 \pi \frac{\sigma_{1}}{r}$
we receive the following difference equation for $\Psi$, interpreting the
order of operators in $\frac{p}{4 \pi \sigma r^{2}}$ as
$\frac{1}{4 \pi \sigma r^{2}}p$
$f(x)\Psi(x)+g(x)[\Psi (x+i) + \Psi (x-i)]=0$
f and g are real functions of x, inside the horizon, and
\begin{equation}
f(x)\Psi(x)+g(x)[\Psi (x+i) - \Psi (x-i)]=0
\end{equation}
outside.
Expanding  the equation for $\Psi$ outside the horizon, taking $x>>1$
(setting $r\ \sim \frac{1}{\chi}$ and $\chi\ \sim \ G^{1/2}\ \rho _{0}^{1/2}$,
we see that $x>>1$ is satisfied if the typical energy scales determining
$\sigma _{0},\ \sigma _{1}\ and \ \rho _{0}$ are $<<$ Planck scale)
and keeping the first nonvanishing contribution only,
 we receive the equation:
\begin{equation}
-\frac{f}{2g} \Psi (x)=i\frac{\partial \Psi}{\partial x}
\end{equation}
It has the form of a Schroedinger equation (x is time-like outside
the horizon). The solution is
$\Psi = Ce^{i \int (-\frac{f}{2g} ) dx}$
where C=constant.
This means that once a bubble passes the horizon it will expand indefinitely,
since  $|\Psi|^{2}$ = constant and therefore
the modulus of the amplitude for the bubble being at
$r=\frac{1}{\chi} +\epsilon$ ($\epsilon > 0 $ is very small) is the same
as the amplitude for the membrane being at r $\rightarrow \ \infty$ with
probability equal 1. Therefore if the wave function
of the  bubble has a tail long enough, so it can get the horizon
$r=\frac{1}{\chi}$, we have the possibility of the formation of
an infinite size bubble i.e. the formation of a universe.
The resulting universe becomes large not by expanding
and displacing an exterior region. This cannot do, since the interior
has negative pressure and the outside zero pressure. Really 
the bubble expands forming a wormhole region that disconnects
from the outside creating a "baby" universe in this case\cite{G}.

\end{document}